\documentclass[11pt]{article}

\usepackage[T1]{fontenc}
\usepackage[margin=1in]{geometry}
\usepackage{amsmath}
\usepackage{amssymb}
\usepackage{graphicx}
\usepackage{booktabs}
\usepackage{tabularx}
\usepackage{multirow}
\usepackage{hyperref}
\usepackage{xcolor}
\usepackage{fancyhdr}
\usepackage{setspace}
\title{Commissioning and clinical outcome assessment of a novel surface-guided radiation therapy (SGRT) system at a C-Arm linear accelerator}

\author{Hui Khee Looe\textsuperscript{1}, Niklas Felix Hendrik Bartner\textsuperscript{1}, Bj\"orn Poppe\textsuperscript{1}, Kay C. Willborn\textsuperscript{1} \\[0.5em]
\small\textsuperscript{1} University Clinic for Medical Radiation Physics and Clinic for Radiation Therapy, \\
\small Medical Campus Pius Hospital, Carl von Ossietzky University, Oldenburg, Germany}
\date{\today}

\begin{document}

\maketitle

\begin{abstract}
\textbf{Background} Surface-guided radiation therapy (SGRT) has become standard-of-care in modern radiotherapy, enabling radiation-dose-free, marker-free patient positioning with submillimeter accuracy. A novel SGRT system, LUNA 3D (LAP, L\"uneburg, Germany) featuring browser-based interface, GPU-accelerated surface reconstruction, high frame rates exceeding 12 Hz, large field-of-view, and virtual laser projection capabilities, has been commissioned and implemented clinically.

\textbf{Purpose} This study reports comprehensive commissioning procedures and the associated results for the LUNA 3D system and evaluates its clinical impact on patient positioning accuracy for breast and pelvic cancer treatments.

\textbf{Methods} Commissioning tests were performed, including temperature drift, reproducibility, translational and rotational shift accuracy, camera occlusion assessment during gantry rotation, radiographic verification against cone-beam computed tomography (CBCT), and End-to-End dosimetric testing. Test results were compared using two reference surfaces: one captured by LUNA 3D (SGRT-reference) and the other derived from the CT external structure (SIM-reference). A clinical evaluation was conducted to compare CBCT-derived positioning corrections from 192 breast and 259 pelvic treatment datasets acquired before and after the clinical implementation of LUNA 3D, employing Welch's two-sample $t$-test and Cohen's $d$ effect size.

\textbf{Results} The temperature drift within the operating temperature is found to be smaller than 0.4 mm for all three axes. Commissioning demonstrated submillimeter accuracy using SGRT-reference (maximum deviations $\leq 0.3$ mm translational, $\leq 0.2^\circ$ rotational). The maximum deviations recorded using SIM-reference are higher (maximum deviations $\leq 0.8$ mm translational, $\leq 0.2^\circ$ rotational), attributable to a systematic bias introduced by the CT-derived reference surface. Radiographic verification showed agreement within 1.0 mm between LUNA 3D and CBCT corrections for both reference surfaces. End-to-End testing yielded CBCT residuals of 0.9--1.3 mm with 1.2\% dosimetric deviation. All performance metrics satisfied ESTRO-ACROP guidelines. Clinical implementation resulted in significant positioning improvements: for breast treatments, 3D translational vector decreased 28.7\% from $7.00 \pm 4.35$ mm to $4.99 \pm 2.75$ mm ($p < 0.001$, Cohen's $d = 0.54$); for pelvic treatments, 3D rotational vector decreased 24.0\% from $2.31 \pm 0.96^\circ$ to $1.76 \pm 0.67^\circ$ ($p < 0.001$, Cohen's $d = 0.66$).

\textbf{Conclusions} LUNA 3D demonstrated high technical accuracy and reproducibility in commissioning tests, meeting international SGRT guidelines. Clinical implementation produced statistically significant and clinically meaningful improvements in patient positioning accuracy with anatomical site-specific benefits. These findings establish LUNA 3D as a reliable SGRT technology that enhances positioning accuracy in routine clinical practice.
\end{abstract}

\section{Introduction}

In recent years, surface-guided radiation therapy (SGRT) has evolved into standard-of-care in modern radiotherapy practices, representing a paradigm shift from conventional laser-based positioning to advanced optical surface imaging for patient setup and treatment monitoring\cite{1,2}. Surface-guided systems utilize structured-light projection and stereoscopic imaging technologies to reconstruct three-dimensional patient surfaces in real-time, enabling radiation-dose-free, marker-free patient positioning guidance with submillimeter spatial accuracy\cite{3}. Beyond static positioning, SGRT systems provide continuous real-time monitoring capabilities and support advanced respiratory motion management techniques, such as deep-inspiration breath-hold (DIBH) for breast irradiation, where cardiac dose reduction is critical\cite{4}.

The clinical advantages of SGRT have been demonstrated across multiple anatomical sites and treatment techniques\cite{5,6}. Numerous studies have reported reductions in setup time following SGRT implementation compared to traditional laser-based workflows\cite{4}. The improved reproducibility of patient positioning achieved through six-degree-of-freedom (6DoF) offset measurements captured by SGRT systems can potentially reduce the frequency and necessity of ionizing radiation-based image guidance\cite{3,7}. The superior positioning accuracy enabled by SGRT could support tighter planning margins, improved target coverage, and enhanced normal tissue sparing\cite{4,5,7}. Studies comparing SGRT with conventional positioning methods have demonstrated that surface-guided alignment yields comparable or enhanced accuracy to three-point marker-based localization, with performance varying by anatomical site\cite{7}.

Recent international guidelines have established comprehensive recommendations for SGRT implementation. The AAPM Task Group 302 report provides detailed guidance on commissioning, clinical use, and quality assurance (QA) for SGRT systems, emphasizing the importance of risk assessment and systematic quality management programs\cite{8}. The ESTRO-ACROP guideline complements these recommendations with European consensus on procurement, workflows, and site-specific QA procedures\cite{1}. Both guidelines also emphasize that SGRT should be implemented as a complementary technology, but not substituting image guidance, particularly for deep-seated targets, where surface position may not reliably correlate with internal anatomy.

Recently, a novel SGRT system, LUNA 3D (LAP, L\"uneburg, Germany), has been introduced to the market. The system features several technological advances including a browser-based user interface enabling remote access and simplified workflow integration, GPU-accelerated parallel computing for high-speed surface reconstruction, large field-of-view coverage, high frame rates exceeding 12 Hz with latency below 250 milliseconds, and a virtual laser concept that projects positioning guidance directly onto the recorded patient surface. These technical specifications position LUNA 3D as a competitive addition to the SGRT market, potentially addressing limitations of earlier-generation systems.

In this work, comprehensive commissioning procedures for the LUNA 3D system installed at a C-arm linear accelerator are reported. The systematic testing approach encompasses reproducibility, translational and rotational shift accuracy, assessment of camera occlusion effects during gantry rotation, radiographic verification against cone-beam computed tomography (CBCT), and End-to-End dosimetric testing. These commissioning activities establish baseline performance characteristics and validate the system's compliance with international QA guidelines. Furthermore, a clinical outcome evaluation comparing positioning accuracy before and after LUNA 3D implementation is presented, providing benchmark data for future investigations.

\section{Materials and Methods}

\subsection{SGRT System -- LUNA 3D}

The LUNA 3D SGRT system (LAP, L\"uneburg, Germany) was installed at an Elekta C-Arm linear accelerator (linac). The installation consists of three camera pods (two cameras per pod) utilizing structured-light projection technology. The system employs multiple projectors that cast a defined pattern onto the patient surface, which is captured by high-resolution cameras from various angles. Through triangulation algorithms, a precise three-dimensional surface model is reconstructed in real-time with a frame rate exceeding 12 Hz and latency below 250 milliseconds. The system features a browser-based user interface with GPU-accelerated parallel computing, providing six-degree-of-freedom (6DoF) tracking of patient positioning. All measurements were performed using the software version 1.2.3.

\begin{figure}[h!]
\centering
\includegraphics[width=\textwidth]{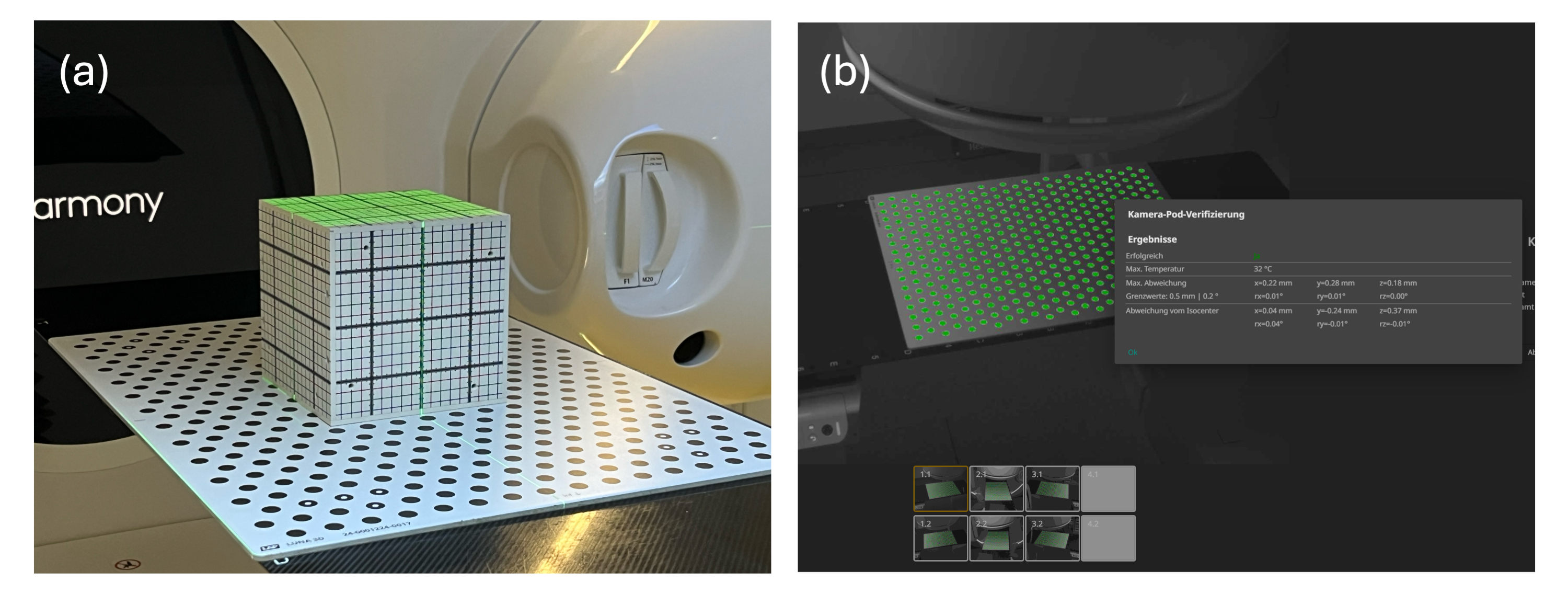}
\caption{(a) Isocenter alignment setup for LUNA 3D system. Geometrical calibration plate with EASY CUBE phantom positioned at the linac isocenter. Both the plate and phantom are aligned with the lateral (cranio-caudal) and longitudinal (left-right) room lasers, while the vertical laser is aligned with the center of the phantom. (b) The daily quality assurance is performed using the geometrical plate alone. LUNA 3D software interface displaying the captured calibration plate position surface and the evaluated accuracy of the cameras' calibrations as well as the isocenter alignment.}
\label{fig:setup}
\end{figure}

The system's isocenter alignment is performed using a geometrical plate with an EASY CUBE phantom (LAP, L\"uneburg, Germany) as shown in Figure~\ref{fig:setup}a. Both the geometrical plate and the phantom are then aligned with the lateral (cranio-caudal) and longitudinal (left-right) room lasers; while the vertical laser is aligned with the center of the phantom. The daily QA is performed only with the geometrical plate as shown in Figure~\ref{fig:setup}b. The plate is also aligned with the room lasers at source-to-surface distance (SSD) of 100 cm. Using the images acquired with the six cameras, the calibration and the isocentric alignment accuracy are evaluated with a tolerance of 0.5 mm and $0.2^\circ$.

\subsection{Test Procedures}

The tests conducted in this work include temperature drift, reproducibility, translational and rotational shift accuracy, impact of camera occlusion, radiographic verification and End-to-End testing. The RUBY Modular Phantom (PTW Dosimetry, Freiburg, Germany) served as the primary phantom for commissioning measurements. This modular phantom features a geometric base design with well-defined edges and surfaces suitable for surface tracking verification. Additionally, the phantom supports defined translational shifts as well as a combination of defined translational and rotational shifts, both with dedicated markings on the phantom's surface.

A CT scan of the phantom was performed at a goSim CT (Siemens Healthineers, Erlangen, Germany) with 1 mm slice thickness. The external structure of the phantom was generated using AI auto-contouring in Elekta ONE Planning (Elekta AB, Stockholm, Sweden). For all tests, except the End-to-End testing (see Section~\ref{sec:e2e}), a treatment plan with isocenter at the center of the phantom, as indicated by markings on the phantom surface was created.

\subsubsection{Temperature Drift}

To measure the temperature-dependent drift of the LUNA 3D system, the calibration plate was placed at the isocenter position. Each of the three camera pods continuously and independently measured the position of the calibration plate. During the measurements, the projectors were turned on and off consecutively for 60 minutes to introduce and remove thermal load. The procedure was repeated twice. The 3D positions, along with the temperatures recorded by the internal built-in sensors, were registered.

\subsubsection{Reproducibility}
\label{sec:reproducibility}

\begin{figure}[h!]
\centering
\includegraphics[width=\textwidth]{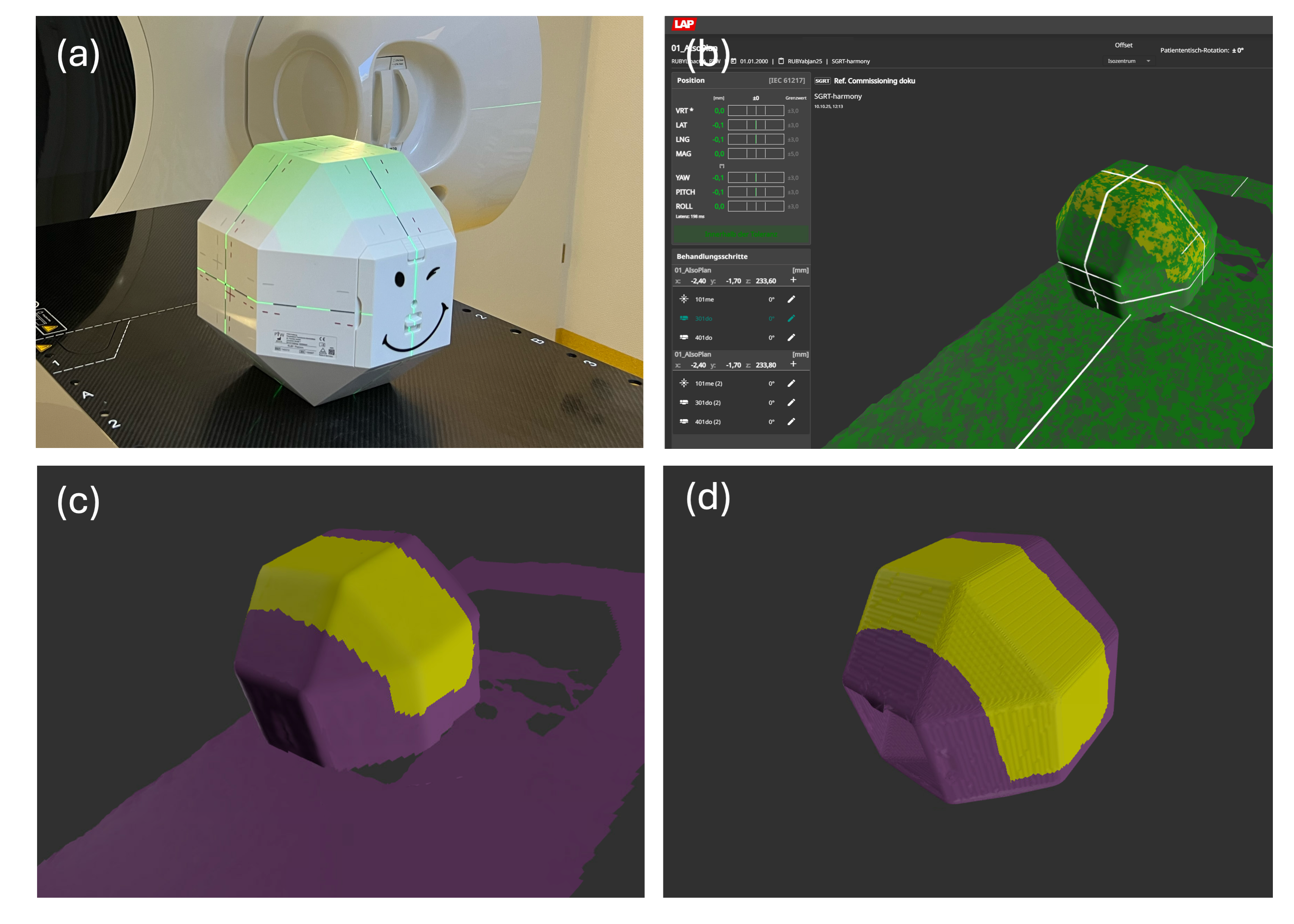}
\caption{Reproducibility test setup with RUBY phantom. (a) The phantom positioned at isocenter using room laser alignment with surface markings. (b) LUNA 3D software interface screenshot showing the live surface (green) and measured 6DoF values. (c) SGRT-reference: reference surface captured by LUNA 3D. (d) SIM-reference: reference surface derived from the external structure of the phantom's CT scan. The ROI defined for tracking (derivation of 6DoF values) is indicated as yellow surface overlay on the reference surface.}
\label{fig:reproducibility}
\end{figure}

The reproducibility measurements were performed by first positioning the RUBY phantom at the isocenter by aligning the surface markings on the phantom with the room lasers. A reference surface was then captured with LUNA 3D (SGRT-reference). Subsequently, the phantom was removed from the couch and repositioned five times. The recorded 6DoF positioning deviations from the reference were recorded and the maximum deviation across all repetitions was documented for each DoF. Figure~\ref{fig:reproducibility} shows (a) the RUBY phantom positioned at the isocenter position and (b) a cropped screenshot of the software interface. 

The same test was repeated using the external structure derived from the phantom's CT scan as the reference surface (SIM-reference). Similarly, the phantom was repositioned five times. The SGRT-reference and the SIM-reference are both are shown in Figure~\ref{fig:reproducibility}c and~\ref{fig:reproducibility}d, respectively, where the region of interest (ROI) is shown as yellow surface.

\subsubsection{Translational and Rotational Shifts}

\begin{figure}[h!]
\centering
\includegraphics[width=\textwidth]{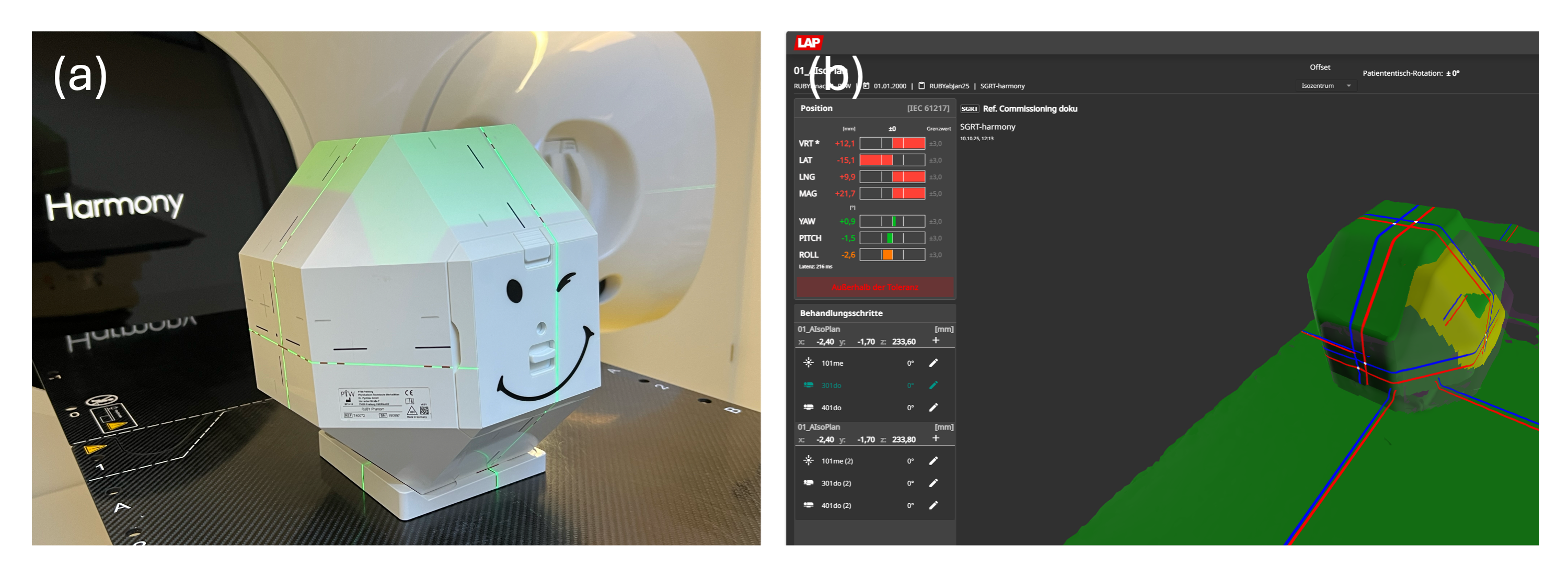}
\caption{Translational and rotational shift accuracy testing setup. (a) RUBY phantom positioned on dedicated tilting base providing predefined translational shifts (12 mm vertical, 10 mm longitudinal, 15 mm lateral) and rotational shifts ($1^\circ$ yaw, $-1.5^\circ$ pitch, $-2.5^\circ$ roll) relative to the isocenter position. (b) cropped LUNA 3D software interface displaying the six 6DoF positioning deviations from the reference surface. The deviations between the fixed red virtual lasers indicating the room isocenters and the blue virtual lasers indicating the isocenter defined in the phantom (treatment plan) also reflect these introduced shifts. In clinical practice, these serve as intuitive visual feedback to the user, whose aim is to align the blue lines to the red lines.}
\label{fig:shifts}
\end{figure}

Using the same reference surfaces as in Section~\ref{sec:reproducibility}, the phantom was repositioned by placing the RUBY phantom on the dedicated tilting base that shifts the phantom by predefined translations and rotations relative to the isocenter position. The setup is shown in Figure~\ref{fig:shifts}a. The procedure was repeated five times. The 6DoF values were recorded (Figure~\ref{fig:shifts}b), where the maximum deviation from the expected values across all repetitions was documented for each DoF.

\subsubsection{Impact of Camera Occlusion}

Using the same setup and reference as in Section~\ref{sec:reproducibility}, the maximum deviation of the measured 6DoF values during gantry rotation from the baseline values at $0^\circ$ with retracted and extended kV X-ray source as well as the kV and MV electronic portal imaging devices (EPID), were recorded, alongside with the gantry angle when this occurred.

\subsubsection{Radiographic Verification}

For the radiographic verification, the RUBY phantom was deliberately positioned with defined translational shifts. The same reference surfaces as in Section~\ref{sec:reproducibility} were used. A CBCT was acquired with the phantom in the shifted position. The LUNA 3D-derived positioning corrections were compared against CBCT-based corrections. Subsequently, the treatment couch was shifted according to CBCT-derived offsets, with residual positioning errors documented to account for mechanical limitations of the couch positioning system.

\subsubsection{End-to-End Dosimetric Testing}
\label{sec:e2e}

\begin{figure}[h!]
\centering
\includegraphics[width=\textwidth]{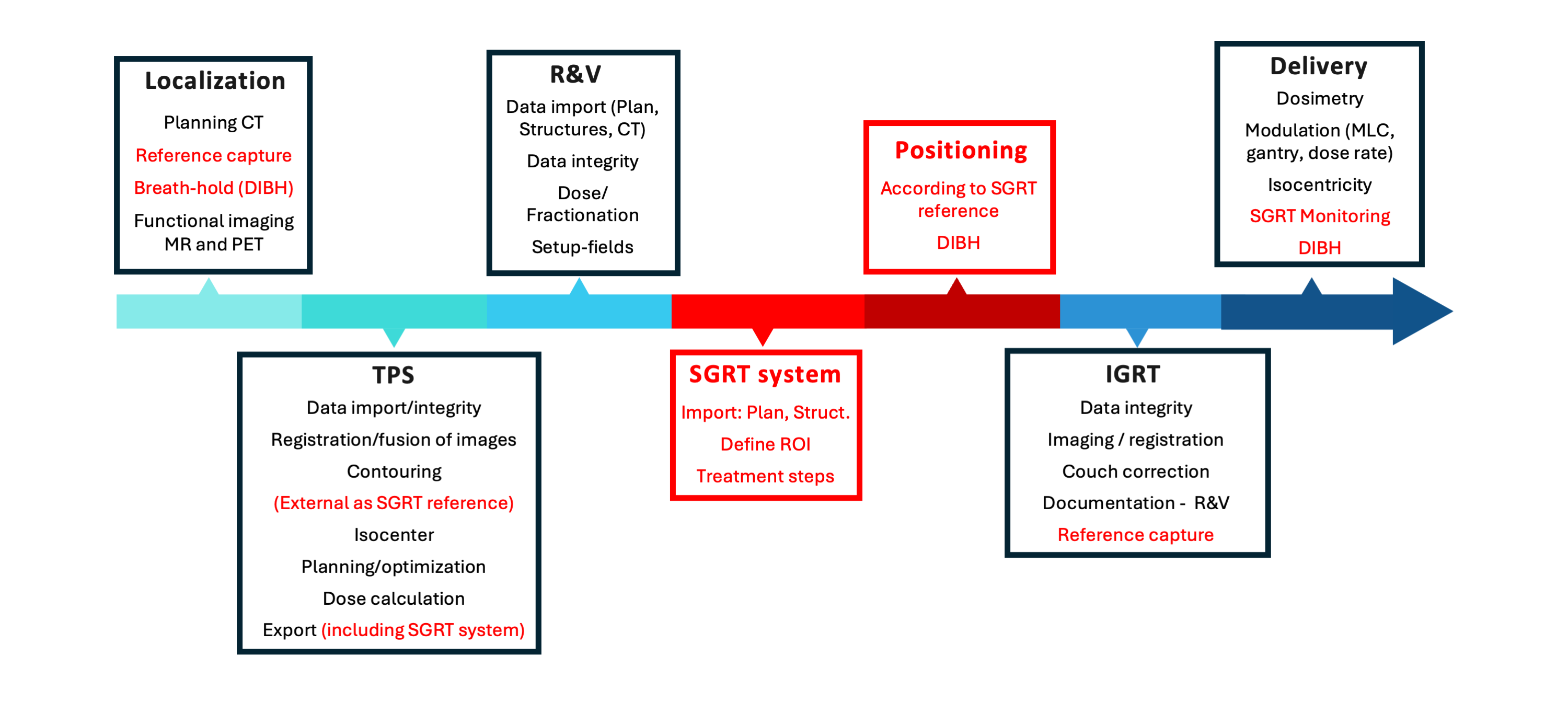}
\caption{End-to-End dosimetric testing workflow incorporating laser-free positioning with LUNA 3D. Schematic representation of the complete workflow from CT simulation through treatment delivery and dosimetric verification. Steps highlighted in red indicate LUNA 3D-specific procedures. The workflow validates the complete laser-free positioning capability from simulation through treatment.}
\label{fig:e2e}
\end{figure}

The procedure for the End-to-End dosimetric testing is depicted in Figure~\ref{fig:e2e}. The procedure was derived based on an existing workflow using the RUBY phantom in combination with the SystemQA insert\cite{9,10}. The test has been adapted to incorporate the laser-free positioning workflow with LUNA 3D (steps in red).

The CT scan of the RUBY phantom with the SystemQA insert was performed as described previously, where the external structure of the phantom was generated using AI auto-contouring as in clinial routine. Subsequently, the CT and structure set were imported into Monaco (version 6.2.3) treatment planning system (TPS) (Elekta AB, Stockholm, Sweden). A treatment plan was created with the isocenter placed at an arbitrary position within the phantom. This ensures that the phantom at the linac cannot be positioned with the help of external markings on the phantom.

The resulted RT-Plan and RT-Structure set were exported to the LUNA 3D server to create a corresponding treatment course with a reference surface derived from the external structure (SIM-reference). The dataset was also exported to the Mosaiq (version 3.1.1.0) Record and Verify (R\&V) system (Elekta AB, Stockholm, Sweden) as well as the linac x-ray volume imaging (XVI) system (Elekta AB, Stockholm, Sweden) for CBCT acquisition. All other preparation steps were carried out as in clinical cases.

At the linac, the phantom was positioned solely using LUNA 3D with the SIM-reference. During the whole process, the room lasers were switched off. Subsequently, a CBCT was acquired to verify the phantom position. Finally, the plan was irradiated, and the dose was measured using a cross-calibrated microDiamond detector (PTW Dosimetry, Freiburg, Germany) placed in the SystemQA insert at the center of the target, which was then compared to the calculated dose.

\subsection{Evaluation of Clinical Performance}

Five months after the clinical implementation of the LUNA 3D system, a retrospective analysis was performed to compare CBCT-derived positioning corrections obtained with LUNA 3D to those acquired prior to its implementation, by recording the CBCT-based translational shifts (lateral, longitudinal, and vertical) and rotational shifts (pitch, roll, and yaw). All shifts were converted to absolute values for analysis.

To provide a comprehensive assessment of overall positioning accuracy, 3D vector magnitudes were calculated for both translational and rotational shifts using the Euclidean distance formula:
\begin{equation}
\text{3D vector} = \sqrt{a^2 + b^2 + c^2}
\end{equation}
where $a$, $b$, and $c$ represent the three orthogonal components (lateral, longitudinal, and vertical for translation; pitch, roll, and yaw for rotation). This metric captures the combined positioning error across all dimensions and provides a clinically relevant measure of overall geometric uncertainty.

Sample sizes for the retrospective analysis comprised 106 CBCT datasets without LUNA and 86 datasets with LUNA for breast treatments, and 143 CBCT datasets without LUNA and 116 datasets with LUNA for pelvic treatments. Descriptive statistics including mean and standard deviation were calculated for each directional component and 3D vector magnitude.

Statistical comparisons between the positioning data with and without LUNA 3D were performed using Welch's two-sample $t$-test with a significance level of $\alpha = 0.05$. Welch's $t$-test was selected as it does not assume equal variances between groups and is robust for comparing independent samples with potentially different population variances. Effect sizes were quantified using Cohen's $d$ to assess the practical significance of observed differences, with interpretation following standard conventions: small effect ($d = 0.2$), medium effect ($d = 0.5$), and large effect ($d = 0.8$).

\section{Results}

\subsection{Temperature Drift}

\begin{figure}[h!]
\centering
\includegraphics[width=\textwidth]{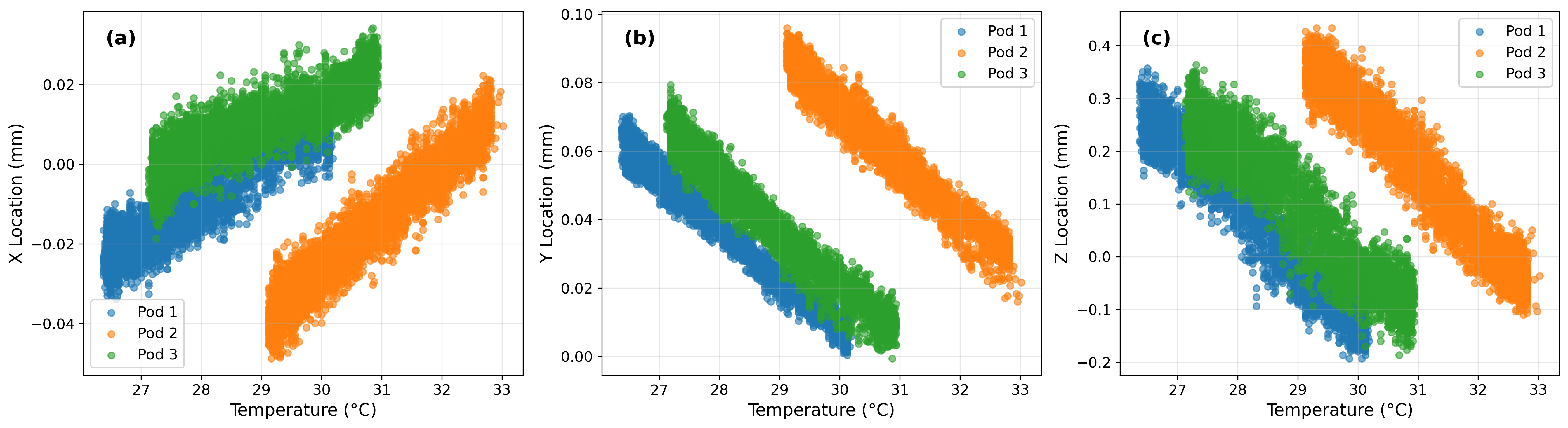}
\caption{Spatial location drift in (a) X-, (b) Y-, and (c) Z-direction measured by each camera pod (Pod 1 to 3) as a function of the corresponding temperature registered by the built-in sensor in the camera pod. All location values are normalized to the first measurement point.}
\label{fig:temperature}
\end{figure}

Figure~\ref{fig:temperature} presents the measured X, Y, and Z positions of the calibration plate by each camera pod as scatter plots against the registered associated temperature of the camera pod. To better assess the effect of thermal drift, all position values were normalized to the first measured values. The thermal drift effects introduced by the projectors occur mainly in the Z direction (distance to camera pod) with a maximum magnitude of 0.4 mm, while the effects are minimal ($\leq 0.1$ mm) in the X and Y directions.

\subsection{Reproducibility}

The LUNA 3D system demonstrated high reproducibility across repeated positioning attempts. When utilizing SGRT-reference, the maximum deviations remained within 0.1 mm for vertical direction, 0.2 mm for longitudinal and lateral directions, and $0.1^\circ$ for all rotational dimensions across five independent measurements.

When using the SIM-reference, the system showed larger maximum deviations reaching $-0.8$ mm vertical, 0.2 mm longitudinal, and 0.4 mm lateral for translational dimensions. Rotational deviations remained within $0.2^\circ$ for all three rotational DoF across five independent measurements. Notably, SIM-reference results revealed a systematic bias of $-0.7$ mm in the vertical direction, while longitudinal and lateral directions showed smaller biases of $+0.2$ mm and $+0.3$ mm, respectively.

\subsection{Translational and Rotational Shifts}
\label{sec:results-shifts}

Using the dedicated tilting base, the RUBY phantom was shifted with known displacements of 12 mm vertical, 10 mm longitudinal, and 15 mm lateral, combined with rotations of $1^\circ$ yaw, $-1.5^\circ$ pitch, and $-2.5^\circ$ roll. Using SGRT-reference, detection accuracy for predefined translational and rotational shifts demonstrated good agreement with expected values with maximum deviations of $-0.2$ mm vertical, 0.1 mm longitudinal, $-0.3$ mm lateral, as well as $-0.2^\circ$ yaw, $0^\circ$ pitch, and $-0.1^\circ$ roll.

When using the SIM-reference, the system demonstrated larger translational deviations. Maximum deviations from expected values reached $-0.8$ mm vertical, 0.3 mm longitudinal, and $-0.5$ mm lateral across five repetitions. The larger deviations, especially in the vertical direction, are consistent with the systematic bias associated with SIM-reference. Rotational accuracy remained high with maximum deviations of $-0.2^\circ$ yaw, $0^\circ$ pitch, and $-0.1^\circ$ roll.

\subsection{Impact of Camera Occlusion}

Using SGRT-reference with all components retracted, maximum deviations remained within 0.1 mm for all translational directions and $0.1^\circ$ for all rotational dimensions across the full $360^\circ$ gantry rotation. With components fully extended (maximum occlusion), identical performance was observed with no detectable differences in measurement accuracy.

Using SIM-reference and components retracted, maximum deviations reached $-0.9$ mm vertical (at $-15^\circ$), $-0.6$ mm longitudinal (at $-85^\circ$), and 1.0 mm lateral (at $-19^\circ$) for translational dimensions. Rotational deviations were all within $0.2^\circ$. With imaging-components fully extended, maximum translational deviations increased slightly to $-1.0$ mm vertical (at $-160^\circ$), 1.0 mm longitudinal (at $-140^\circ$), and $-1.2$ mm lateral (at $-161^\circ$). Rotational deviations increased slightly to $0.3^\circ$.

\subsection{Radiographic Verification}
\label{sec:results-radiographic}

Table~\ref{tab:radiographic} shows the comparison of LUNA 3D and CBCT-based corrections for using both SGRT-reference and SIM-reference. The RUBY phantom was positioned with known translational shifts of $-18$ mm vertical, $-14$ mm longitudinal, and 25 mm lateral. 

With SGRT-reference, the maximum deviations between the corrections registered by LUNA 3D and the expected values are 0.7 mm vertical, $-0.5$ mm longitudinal and $-1$ mm lateral. The corresponding differences between CBCT-based corrections and the expected values are higher than in the results reported in Section~\ref{sec:results-shifts} as the results here were further influenced by additional uncertainty components, including kV and laser isocenter congruence, CT-CBCT registration and isocenter placement in treatment plan. Nevertheless, the maximum difference to the expected values is still within 1 mm.. After the couch correction, the maximum couch residue as indicated on the linac console was $\pm 0.3$ mm. The maximum difference between the remaining LUNA 3D corrections and couch residue values after couch correction amount to 0.6 mm vertical, $-0.4$ mm longitudinal and $-1.3$ mm lateral, consistent with the maximum deviations between LUNA 3D and CBCT-based corrections before the couch correction was executed. The small difference within 0.3 mm is attributable to mechanical limitations of the treatment couch positioning system.

Using the SIM-reference, the maximum deviations between LUNA 3D-derived corrections and expected shift values were $-0.3$ mm vertical, $-0.3$ mm longitudinal, and $-0.5$ mm lateral. After the couch correction, the maximum couch residue as indicated on the linac console was $-0.4$ mm. Following couch correction, the remaining LUNA 3D-indicated deviations and the couch residuals resulted in maximum total difference of $-0.6$ mm vertical, 0.3 mm longitudinal, and $-0.7$ mm lateral. Notably, these differences using SIM-reference are smaller than those observed with SGRT-reference despite the reported systematic bias, suggesting complex interplay between the additional uncertainty components as mentioned earlier and the used reference surface. 

\begin{table}[h]
\centering
\caption{Radiographic verification comparing LUNA 3D and CBCT-based positioning corrections. The RUBY phantom was positioned with predefined shifts of -18.0 mm vertical, -14.0 mm longitudinal, and 25.0 mm lateral. Results are shown for tests (a) using SGRT-reference and (b) using SIM-reference across three repetitions. LUNA 3D and CBCT corrections are shown before couch correction, with remaining LUNA 3D deviations and couch residuals documented after couch correction. Maximum deviations represent the largest discrepancies across all repetitions. All values in mm.}
\label{tab:radiographic}
\textbf{(a) SGRT-reference}

\begin{tabular}{llccc|lccc}
\hline
\multicolumn{5}{c|}{\textbf{Before couch correction}} & \multicolumn{4}{c}{\textbf{After couch correction}} \\
\textbf{Repetition} & & \textbf{Vert} & \textbf{Long} & \textbf{Lat} & & \textbf{Vert} & \textbf{Long} & \textbf{Lat} \\
\hline
1 & LUNA 3D & $-18.1$ & $-13.9$ & $24.7$ & LUNA 3D & $\phantom{-}0.1$ & $\phantom{-}0.2$ & $-0.9$ \\
  & CBCT & $-18.6$ & $-13.6$ & $25.5$ & Couch residue & $-0.2$ & $\phantom{-}0.3$ & $\phantom{-}0.1$ \\
2 & LUNA 3D & $-18.2$ & $-13.9$ & $24.8$ & LUNA 3D & $\phantom{-}0.4$ & $\phantom{-}0.1$ & $-0.9$ \\
  & CBCT & $-18.9$ & $-14.1$ & $25.7$ & Couch residue & $\phantom{-}0.2$ & $-0.2$ & $\phantom{-}0.0$ \\
3 & LUNA 3D & $-17.9$ & $-14.2$ & $24.6$ & LUNA 3D & $\phantom{-}0.5$ & $-0.7$ & $-1.5$ \\
  & CBCT & $-18.4$ & $-13.7$ & $25.6$ & Couch residue & $-0.1$ & $-0.3$ & $-0.2$ \\
\hline
  & max dev. & $\phantom{-}0.7$ & $-0.5$ & $-1.0$ & max dev. & $\phantom{-}0.6$ & $-0.4$ & $-1.3$ \\
\hline
\end{tabular}

\vspace{1em}

\textbf{(b) SIM-reference}

\begin{tabular}{llccc|lccc}
\hline
\multicolumn{5}{c|}{\textbf{Before couch correction}} & \multicolumn{4}{c}{\textbf{After couch correction}} \\
\textbf{Repetition} & & \textbf{Vert} & \textbf{Long} & \textbf{Lat} & & \textbf{Vert} & \textbf{Long} & \textbf{Lat} \\
\hline
1 & LUNA 3D & $-18.5$ & $-13.9$ & $25.4$ & LUNA 3D & $-0.4$ & $\phantom{-}0.3$ & $-0.8$ \\
  & CBCT & $-18.4$ & $-13.7$ & $25.6$ & Couch residue & $\phantom{-}0.0$ & $\phantom{-}0.3$ & $-0.1$ \\
2 & LUNA 3D & $-18.6$ & $-13.6$ & $25.2$ & LUNA 3D & $-0.6$ & $\phantom{-}0.4$ & $-0.8$ \\
  & CBCT & $-18.4$ & $-13.7$ & $25.7$ & Couch residue & $-0.1$ & $\phantom{-}0.1$ & $-0.3$ \\
3 & LUNA 3D & $-18.6$ & $-13.5$ & $24.9$ & LUNA 3D & $-0.6$ & $-0.6$ & $-0.6$ \\
  & CBCT & $-18.3$ & $-13.2$ & $25.3$ & Couch residue & $\phantom{-}0.0$ & $-0.4$ & $-0.1$ \\
\hline
  & max dev. & $-0.3$ & $-0.3$ & $-0.5$ & max dev. & $-0.6$ & $\phantom{-}0.3$ & $-0.7$ \\
\hline
\end{tabular}
\end{table}

\subsection{End-to-End Dosimetric Testing}

\begin{figure}[h!]
\centering
\includegraphics[width=\textwidth]{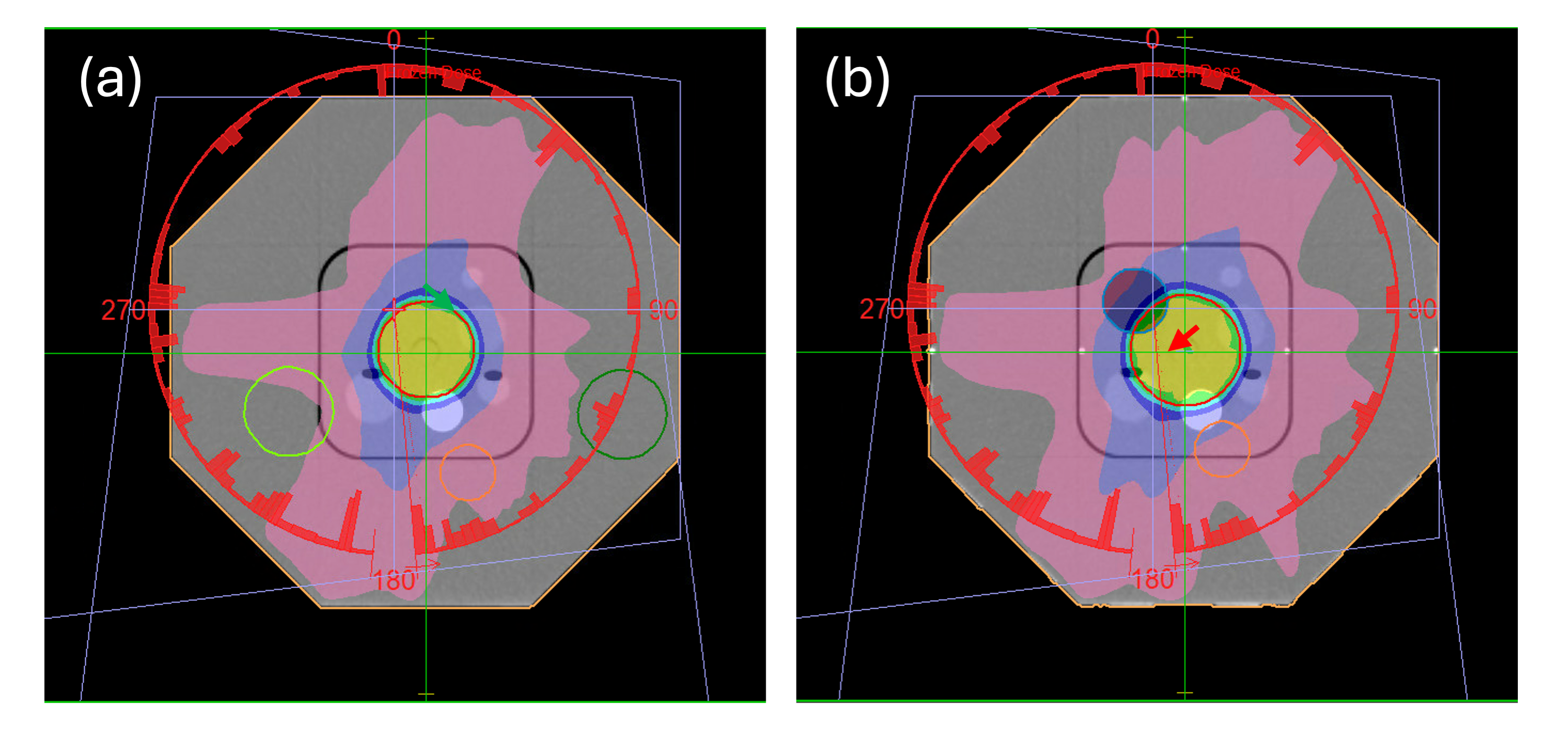}
\caption{End-to-End testing dosimetric verification. (a) Transversal CT slice of the RUBY phantom showing the treatment isocenter (green arrow) positioned at an arbitrary location within the phantom to prevent external marker-based positioning. (b) Target volume centered at the measurement position, where the microDiamond detector is placed in the SystemQA insert at the center of the phantom indicated by red arrow.}
\label{fig:e2e_dosimetry}
\end{figure}

Figure~\ref{fig:e2e_dosimetry}a shows the transversal slice of the RUBY phantom, in which the isocenter has been defined as indicated by the green arrow. The dose was measured at the center of the phantom, at which the target was also centered as indicated by the red arrow in Figure~\ref{fig:e2e_dosimetry}b. The CBCT-based corrections after the phantom was positioned solely with LUNA 3D amounted to 0.9 mm vertical, $-0.6$ mm longitudinal and $-1.3$ mm lateral. The values are comparable to the results of radiographic verification presented in the previous Section~\ref{sec:results-radiographic} using SIM-reference. The measured dose using the microDiamond detector (3.106 Gy) agreed to the calculated dose (3.142 Gy) with a deviation of $-1.2\%$.

\subsection{Evaluation of Clinical Performance}

\begin{figure}[h!]
\centering
\includegraphics[width=\textwidth]{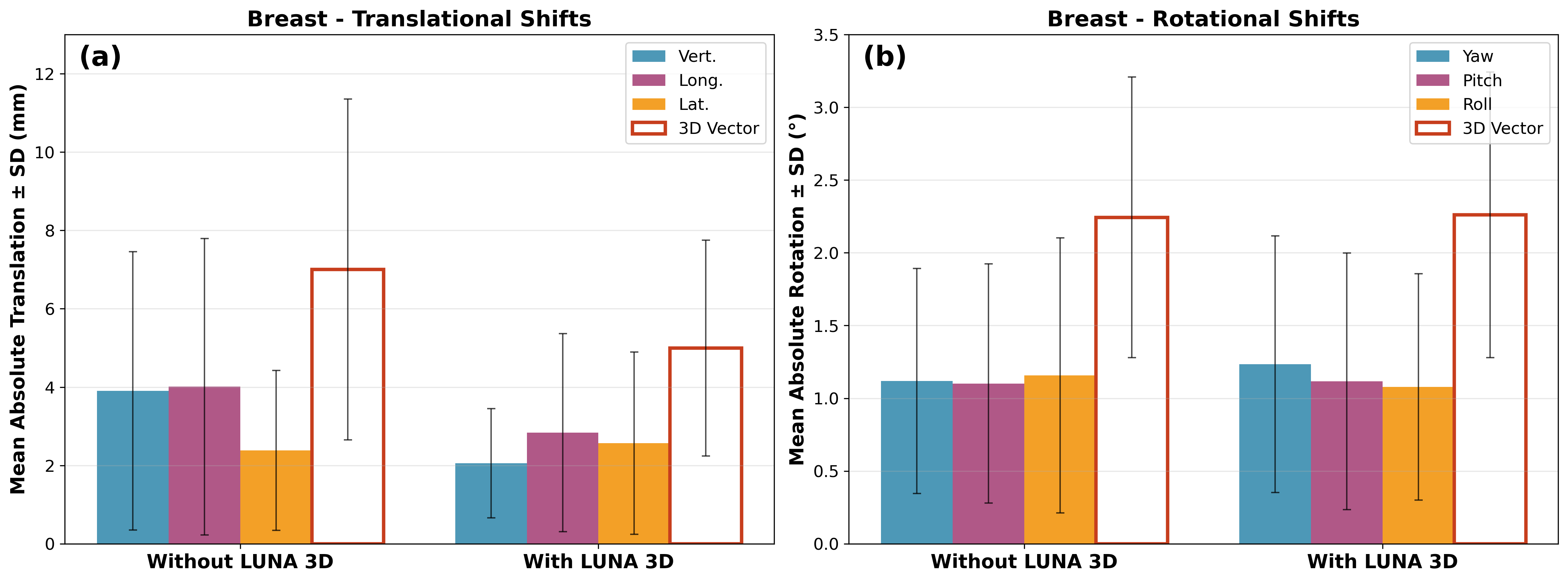}\\
\includegraphics[width=\textwidth]{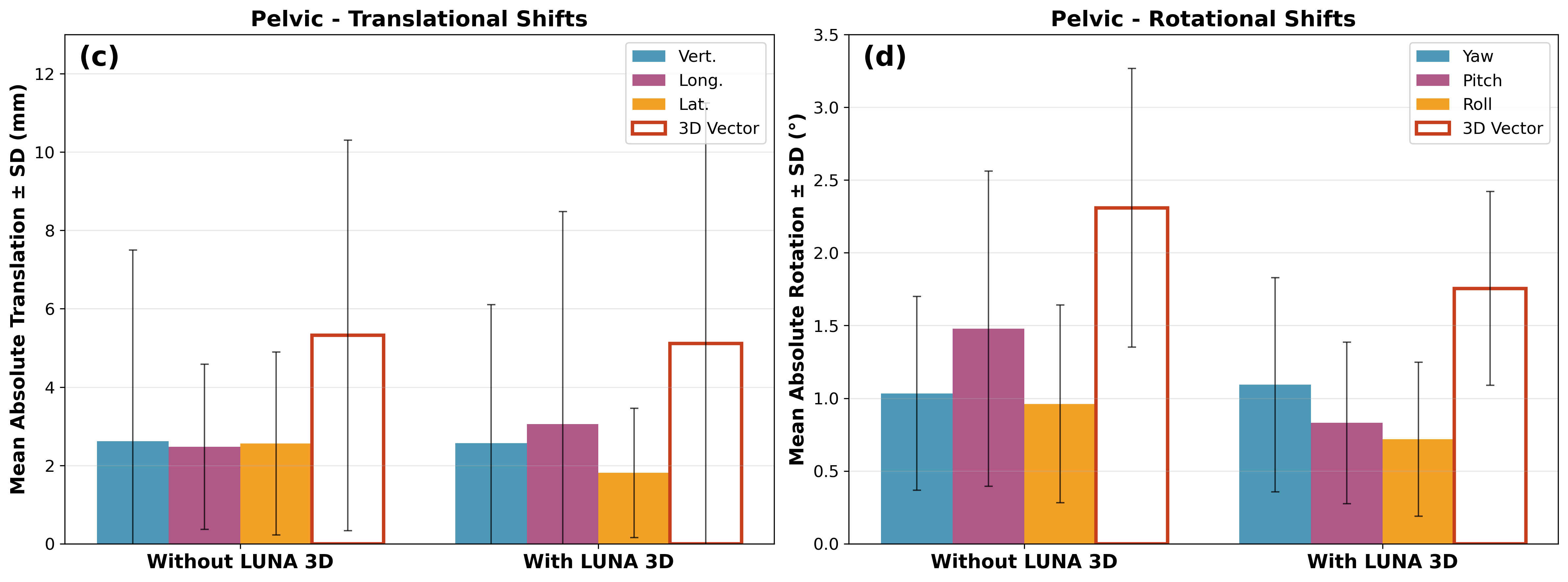}
\caption{Positioning accuracy with and without LUNA 3D surface guidance. Bar plots comparing mean absolute positioning errors ($\pm$ SD) for (a, b) breast and (c, d) pelvic treatments. (a, c) translational shifts in millimeters and (b, d) rotational shifts in degrees are shown separately. The individual directional components are shown in shaded grey bars, while the 3D vector magnitudes (calculated as Euclidean distance: $\sqrt{a^2+b^2+c^2}$), are shown as hollow bars. For breast treatments, LUNA significantly reduced translational 3D vector errors by 28.7\% ($7.00 \pm 4.35$ mm vs. $4.99 \pm 2.75$ mm, $p < 0.001$, Cohen's $d = 0.54$) but had no significant effect on rotational errors ($p = 0.897$). For pelvic, LUNA significantly reduced rotational 3D vector errors by 24.0\% ($2.31 \pm 0.96^\circ$ vs. $1.76 \pm 0.67^\circ$, $p < 0.001$, Cohen's $d = 0.66$) but had no significant effect on translational errors ($p = 0.771$).}
\label{fig:clinical}
\end{figure}

Retrospective analysis of CBCT positioning corrections revealed measurable improvements in positioning accuracy following LUNA 3D implementation for both breast and pelvic cancer treatments as presented in Figure~\ref{fig:clinical}. 

For breast cancer patients, significant improvements were observed in overall translational positioning accuracy. The 3D translational vector decreased significantly from $7.00 \pm 4.35$ mm to $4.99 \pm 2.75$ mm, representing a 28.7\% improvement ($t = 3.89$, $p < 0.001$, Cohen's $d = 0.54$). This medium effect size indicates a clinically meaningful improvement in translational positioning accuracy. Individual directional components showed vertical deviations decreasing from $3.91 \pm 3.55$ mm to $2.06 \pm 1.39$ mm (47.3\% reduction), longitudinal deviations reducing from $4.01 \pm 3.78$ mm to $2.84 \pm 2.52$ mm (29.2\% reduction), and lateral deviations increasing slightly from $2.39 \pm 2.04$ mm to $2.57 \pm 2.32$ mm. Rotational positioning for breast treatments remained stable, with the 3D rotational vector showing no significant change from $2.24 \pm 0.96^\circ$ to $2.26 \pm 0.98^\circ$ ($t = -0.13$, $p = 0.897$, Cohen's $d = -0.02$). Individual rotational components demonstrated yaw increasing from  $1.12 \pm 0.77^\circ$ to $1.23 \pm 0.88^\circ$, pitch increasing marginally from $1.10 \pm 0.82^\circ$ to $1.12 \pm 0.88^\circ$, and roll decreasing from $1.16 \pm1 0.94^\circ$ to $1.08 \pm 0.78^\circ$.

For pelvic cancer patients, LUNA 3D implementation also resulted in significant improvements in rotational positioning accuracy. The 3D rotational vector decreased from $2.31 \pm 0.96^\circ$ to $1.76 \pm 0.67^\circ$, representing a 24.0\% improvement ($t = 5.47$, $p < 0.001$, Cohen's $d = 0.66$). This medium-to-large effect size indicates a clinically meaningful improvement. Individual rotational components showed pitch decreasing substantially from $1.48 \pm 1.08^\circ$ to $0.83 \pm 0.56^\circ$ (43.8\% reduction), roll reducing from $0.96 \pm 0.68^\circ$ to $0.72 \pm 0.53^\circ$ (25.2\% reduction), and yaw increasing slightly from $1.03 \pm 0.67^\circ$ to $1.09 \pm 0.74^\circ$. Translational positioning for pelvic treatments showed minimal change, with the 3D translational vector decreasing non-significantly from $5.32 \pm 4.98$ mm to $5.11 \pm 6.14$ mm ($t = 0.29$, $p = 0.771$, Cohen's $d = 0.04$). Individual directional components demonstrated lateral positioning decreasing from $2.56 \pm 2.33$ mm to $1.81 \pm 1.65$ mm (29.3\% reduction), vertical positioning reducing slightly from $2.62 \pm 4.88$ mm to $2.57 \pm 3.54$ mm, and longitudinal positioning increasing from $2.48 \pm 2.11$ mm to $3.05 \pm 5.43$ mm. 

These findings demonstrate that surface-guided radiation therapy with LUNA 3D enhanced initial patient positioning accuracy in routine clinical workflows, with statistically significant and clinically meaningful improvements especially observed for breast translational and pelvic rotational positioning.

\section{Discussion}

The commissioning and clinical outcome evaluation of the LUNA 3D SGRT system demonstrated high technical accuracy and anatomical site-specific improvements in patient positioning. The system achieved accuracy in phantom testing that fulfils international SGRT guidelines. Clinical implementation revealed statistically significant improvements of clinical outcomes.

\subsection{Technical Performance}

The thermal drift effects introduced by the projectors occur predominantly in the Z direction (distance to camera pod) with a maximum magnitude of 0.4 mm, while remaining minimal ($\leq 0.1$ mm) in the X and Y directions. The LUNA 3D system achieved high reproducibility testing with SGRT-reference, with maximum deviations of 0.1 mm vertically, 0.2 mm in longitudinal and lateral directions, and $0.1^\circ$ in all rotational dimensions across five independent measurements. When using SIM-reference, the larger deviations in the translational shifts of 0.8 mm vertical, 0.2 mm longitudinal, 0.4 mm lateral are associated with a systematic bias attributable to additional uncertainties related to the differences in the surfaces of different origins (SGRT and CT), as well as pixel discretization in CT and its effect on the surface delineations in CT. However, the maximum rotational deviation remained small at $0.2^\circ$.

The translational and rotational shift accuracy tests further validated the system's precision, with maximum deviations of 0.3 mm translational and $0.2^\circ$ rotational across known phantom displacements. Consistent with the findings in reproducibility test, the maximum translational deviations reached $-0.8$ mm, while rotational accuracy remained identical at $-0.2^\circ$, indicating the same observations of increased deviations when SIM-reference was used.

The impact of camera occlusion during gantry rotation was minimal with SGRT-reference, with deviations remaining at or below 0.1 mm and $0.1^\circ$ throughout the full $360^\circ$ rotation arc with imaging-components retracted or extended. In contrast, the use of SIM-reference showed larger gantry angle-dependent influences, with maximum translational deviations of 1.0 to 1.2 mm occurring at specific angles for a short moment.

The radiographic verification tests demonstrated good agreement between LUNA 3D-derived corrections and CBCT-based corrections, with maximum differences of 0.7 mm vertically, $-0.5$ mm longitudinally, and $-1.0$ mm laterally before couch correction. These discrepancies are within expected ranges when considering the cumulative uncertainties from kV-laser isocenter congruence as well as CT-CBCT registration algorithms. After couch corrections were applied, residual couch positioning errors of up to 0.3 mm were observed, admitting mechanical couch positioning tolerances, differing by up to 1.3 mm (lateral) from LUNA 3D measured residues. Radiographic verification tests using SIM-reference reflected similar, or slightly better results, where the maximum deviations are all within 1 mm.

The End-to-End testing, which utilized a laser-free positioning workflow based entirely on LUNA 3D surface guidance, yielded CBCT-derived residual positioning errors of 0.9 mm vertically, 0.6 mm longitudinally, and 1.3 mm laterally. The measured dose deviated from the calculated dose by only 1.2\%, demonstrating that surface-guided positioning without traditional laser alignment can achieve clinically acceptable accuracy for radiation delivery.

These results align with the ESTRO-ACROP guidelines for SGRT systems, which recommend maximum deviations of $\leq 1$ mm for translations and $\leq 1^\circ$ for rotations during commissioning testing. The observed performance is comparable to established SGRT systems documented in recent literature, confirming that LUNA 3D represents a reliable addition to available surface guidance technologies.

\subsection{Clinical Outcomes}

The retrospective analysis of clinical positioning data revealed substantial and statistically significant improvements in setup accuracy following LUNA 3D implementation. Statistical analysis identified two highly significant improvements (both $p < 0.001$ with medium effect sizes): breast translational positioning (28.7\% reduction in 3D vector magnitude) and pelvic rotational positioning (24.0\% reduction in 3D vector magnitude).

For breast cancer treatments, improvement was driven primarily by enhanced vertical and longitudinal positioning, which are the dominant contributors to setup uncertainty in breast treatments. These positioning improvements would impact target coverage and organ sparing, as well as anterior-posterior alignment\cite{4,6}. Rotational positioning for breast treatments remained stable, with the 3D rotational vector showing no significant change. Previous studies have similarly reported that SGRT systems demonstrate variable rotational performance for breast treatments depending on region of interest selection, anatomical contour stability, and tissue deformability\cite{5,6}.

For pelvic cancer treatments, the rotational improvement was driven primarily by substantial reductions in pitch (43.8\% from 1.48° to 0.83°) and roll (25.2\% from 0.96° to 0.72°) rotations. The substantial pitch improvement suggests that LUNA 3D's surface-based alignment effectively captures and corrects pelvic tilt variations that may be less apparent using conventional laser-based positioning.

Recent literature supports these findings. Jeong et al. analyzed 184 breast cancer patients comparing marker-less SGRT with traditional skin marks, reporting mean 3D vector errors of 5.2 mm with SGRT\cite{11}. These values are consistent with LUNA 3D performance observed in the present study (3D vector = 4.99 mm). Svestad et al. conducted a randomized crossover trial of 25 breast cancer patients, reporting 3D vector errors of 5.1 mm with SGRT versus 5.3 mm with skin marks, demonstrating statistical equivalence\cite{12}. Rigley et al. evaluated tattoo-free SGRT positioning in 43 breast patients, demonstrating significantly improved accuracy compared to conventional methods (4.7 mm versus 5.2 mm, $p = 0.04$), with more pronounced benefits for DIBH treatments (4.5 mm versus 7.6 mm, $p < 0.001$)\cite{13}. Kang et al. reported mixed directional performance with SGRT showing improved vertical accuracy, but larger longitudinal errors compared to skin marks\cite{14}. The clinical performance of LUNA 3D aligns with these recent comparative evaluations of SGRT systems, where the present study extends beyond breast-focused literature by evaluating pelvic treatments, revealing differential effectiveness across anatomical sites and positioning dimensions.

The radiation dose-free nature of SGRT is particularly advantageous for clinical scenarios requiring frequent positioning verification, where cumulative imaging dose becomes clinically relevant. The real-time continuous monitoring capability of SGRT provides an additional safety layer beyond conventional image-guided radiotherapy, where the continuous surveillance enables immediate detection of intra-fraction motion, supporting advanced treatment techniques such as DIBH for breast radiotherapy, respiratory gating, and open-face mask treatments.

The current results also highlight that SGRT should be viewed as complementary to, rather than a complete replacement for, imaging-based position verification. For treatment sites where internal target position may not correlate reliably with external surface anatomy or lung tumors subject to organ motion, combined SGRT and CBCT workflows remain the standard of care. The LUNA 3D system's ability to reduce but not eliminate the need for imaging verification aligns with current clinical guidelines recommending hybrid positioning strategies that leverage the strengths of both SGRT and IGRT\cite{1,8}.

\subsection{Limitations and Future Directions}

The clinical evaluation is at the current stage limited to breast and pelvic cancer patients, precluding definitive conclusions about system performance for other anatomical sites such as head and neck, thorax, or extremity treatments. Additionally, the retrospective nature of the clinical data analysis introduces potential confounding variables, including inter-operator variability in patient positioning technique and differences in patient characteristics between the pre- and post-LUNA 3D implementation cohorts. The sample sizes of CBCT datasets per cohort, while adequate for initial clinical outcome evaluation, are modest compared to large multi-institutional studies. Larger prospective studies would provide more robust estimates of SGRT performance and enable subgroup analyses by patient characteristics, treatment technique, and anatomical site.

Assessment of treatment efficiency metrics was not reported in this study as the retrospective analysis was conducted post-clinical implementation of LUNA 3D. The lack of data of setup time in the pre-implementation period does not support reliable comparison. Although not quantitatively measured in this study, radiation therapists consistently reported perceived workflow improvements and reduced setup times with LUNA 3D. In their setup time analysis, Kang et al. revealed efficiency gains with SGRT (5.2 minutes versus 5.5 minutes)\cite{14}, consistent with Jeong et al.'s findings of reduced positioning times (261 seconds versus 281 seconds)\cite{11}. It is suggested that progressive workflow improvement over time after SGRT implementation could result in further efficiency improvement.

The LUNA 3D software version tested at the time of commissioning offered only limited interoperability with the linear accelerator control system, precluding comprehensive evaluation of its dynamic performance, including beam-hold performance and lag time measurements between detected motion and beam trigger response. Following software updates that enable linear accelerator interoperability, a comprehensive evaluation of LUNA 3D's dynamic performance will be conducted and reported in a follow-up study.

\section{Conclusion}

The commissioning and initial clinical evaluation of the LUNA 3D surface-guided radiation therapy system demonstrated high accuracy, stability, and reproducibility under both phantom testing conditions and routine clinical use. The system achieved the recommended accuracy and precision across all tested configurations, meeting international guidelines for SGRT acceptance and QA.

Clinical implementation of LUNA 3D resulted in measurable improvements in patient positioning accuracy, with the mean 3D translational vector decreased by 18.0\%, while the mean 3D rotational vector improved by 11.6\% combined across both anatomical sites analyzed in this work. The magnitude of improvement varied by anatomical site, with pelvic treatments showing consistent benefits in rotational positioning, while breast treatments demonstrated significant translational improvements.

This study provides initial benchmark data for centers implementing the LUNA 3D system and demonstrates that modern SGRT technology can enhance positioning accuracy in routine clinical practice. The findings contribute to the growing body of evidence supporting SGRT as a standard component of modern radiotherapy quality management.

\section*{Acknowledgements}

The authors gratefully acknowledge the clinical radiation therapist team for their dedication and effort in implementing the LUNA 3D clinical workflows. The authors also wish to thank (in alphabetical order) Michelle B\"ock, Jens Gauthier, Daniel Kayser, Ralf M\"uller-Polyzou, Sascha Neber, Joshua Norton, Raphael Schmidt, Laura Schwarz, Thomas Speck, and Wolfgang von Stubbendorf from LAP GmbH for their continuous support and valuable discussions during the clinical implementation of the LUNA 3D system in our institution.

\section*{Conflicts of Interest}

The authors declare that they have no conflicts of interest or financial interests to disclose. The LUNA 3D system evaluated in this study was independently purchased for routine clinical use.

\section*{Data Availability}

Authors will share data upon request to the corresponding author.

\end{document}